\documentclass[aps,twocolumn]{revtex4}%
\usepackage{amsfonts}
\usepackage{amsmath}
\usepackage{amssymb}
\usepackage{graphicx}
\usepackage{float}%
\providecommand{\U}[1]{\protect\rule{.1in}{.1in}}

\begin{document}
\title{The geometry of von-Neumann's pre-measurement and weak values}
\author{Augusto C\'{e}sar Lobo}
\affiliation{Departamento de F\'{\i}sica - Instituto de Ci\^{e}ncias Exatas e
Biol\'{o}gicas - Universidade Federal de Ouro Preto, CEP 35400-000, Ouro Preto
- Minas Gerais, Brazil}
\author{Clyffe de Assis Ribeiro}
\affiliation{Departamento de F\'{\i}sica, Instituto de Ci\^{e}ncias Exatas, Universidade
Federal de Minas Gerais, CP 702, CEP 30161-970, Belo Horizonte, Minas Gerais, Brazil}
\begin{abstract}
We have carried out a review of a paper from Tamate et \textit{al}, conducting
a deeper study of the geometric concepts they introduced in their paper,
clarifying some of their results and calculations and advancing a step futher
in their geometrization program for an understanding of the structure of
von-Neumann's pre-measurement and weak values.

\end{abstract}
\date{\today}
\maketitle



\section{Introduction}

The concept of a \textit{weak value }of a quantum mechanical system was
introduced in 1988 by Aharonov, Albert and Vaidman \cite{aharonov1988}. It was
built on a \textit{time symmetrical model }for quantum mechanics previously
introduced by Aharonov, Bergmann and Lebowitz in 1964 \cite{aharonov1964}. In
this model, non-local time boundary conditions are used, since the description
of the state of a physical system between two quantum mechanical measurements
is made by pre and post-selection of the states. The authors developed the so
called \textit{ABL Rule} for the transition probabilities within this
scenario, so this is why it is also known as the \textit{two state formalism}
for quantum mechanics \cite{aharonov2007}. The \textit{weak value} of an
observable can be considered as a generalization of the usual expectation
value of a quantum observable, but differently from this, it takes values in
the \textit{complex plane} in general \cite{jozsa2007,augusto2009}. The weak
value concept has shown a plethora of theoretical and experimental
applications. The issue of \textit{quantum counterfactuality}, for instance,
seems to be particularly less paradoxical when analyzed in terms of weak
values \cite{Aharonov2002hardy,Elitzur1993}. In \textit{quantum metrology},
the amplification of tiny effects in quantum mechanics has spawn some recent
impressive results as the observation of the spin Hall effect of light
\cite{Hosten2008}. For a recent review on weak values, see \cite{shikano2011}.

In our present work, we elaborate on a previous paper of Tamate \textit{et al
}where the authors introduce a very interesting geometric interpretation of
the von Neumann pre-measurement and weak values which are closely related
concepts. We conduct a review of their work, making the geometric structures
more mathematically precise and advancing further in this geometrization
programme. We also clarify some calculations and results from their original paper.

In the next section we review the von Neumann ideal pre-measurement formalism
mostly to introduce our notation. In section III, we review Tamate \textit{et
al's} geometric description of the interaction of a system with a discrete
measuring system in a deeper mathematical manner based on the geometry of
quantum mechanics developed by Berry and Aharonov-Anandan among many others
dating back to the eighties \cite{berry1984,aharonov1987}. In section IV, we
discuss their extension to infinite dimensional measuring systems with
continuous indexed basis. We clarify the geometric content of a derivation of
the intrinsic phase between two infinitesimally nearby states in the measured
subsystem induced by an ideal von Neumann pre-measurement. The shift in
position due to the instantaneous interaction with a measuring subsystem is
shown to be proportional to the expectation value of the arbitrary observable
$\hat{O}$ that is being measured in the first subsystem. We show how to
conduct this derivation through a simple but deeper analysis of the
geometrical structures involved. We also extend their calculation of the
position shift in the measuring apparatus for initial states that results
explicitly in a \textit{non-null} imaginary part of the weak value. For the
case of a single qubit this leads to a trivial geometric interpretation of
this complex weak value. Finally, in section V, we address some concluding
remarks and set stage for further work.

\section{The von Neumann pre-measurement model.}

We discuss von Neumann's model for a pre-measurement \cite{neumann1955} where
the measuring apparatus is also considered as a quantum system. Let
$W=W_{S}\otimes W_{M}$ be the state vector space of the system formed by the
subsystem $W_{S}$ and the \textit{measuring }subsystem $W_{M}$. We will also
assume that the measured system is a discrete quantum variable of $W_{S}$
defined by the observable $\hat{O}=|o_{i}\rangle o_{i}\langle o^{i}|$ (the sum
convention will be used hereinafter). The measuring subsystem will be
considered as a structureless (no spin or internal variables) quantum
mechanical particle in one dimension. (In the next section we will consider
discrete measuring systems.) Thus, we can choose as a basis for the vector
state space $W_{M}$ either one of the usual eigenstates of position or
momentum $\{|q(x)\rangle\}$ or $\{|p(x)\rangle\}$. It is important to note
here that we use a slightly different notation than usual (for reasons that
will soon become evident) in the sense that we distinguish between the
\textquotedblleft type\textquotedblright\ of the eigenvector ($q$ or $p$) from
the actual $x$ eigenvalue \cite{augusto2009}. For instance, we write:
\begin{equation}
\hat{Q}|q(x)\rangle=x|q(x)\rangle\qquad\text{and}\qquad\hat{P}|p(x)\rangle
=x|p(x)\rangle.
\end{equation}
(instead of $\hat{Q}|q\rangle=q|q\rangle$ and $\hat{P}|p\rangle=p|p\rangle$ as
commonly written) where $\hat{Q}$ and $\hat{P}$ are the position and momentum
observables subject to the well known Heisenberg relation: $[\hat{Q},\hat
{P}]=i\hat{I}$ (hereinafter, $\hbar=1$ units will be used). With this
non-standard notation, the completeness relation, normalization and the
overlapping between these bases can be written respectively as:
\[
\int\limits_{-\infty}^{+\infty}|q(x)\rangle\langle q(x)|dx=\int
\limits_{-\infty}^{+\infty}|p(x)\rangle\langle p(x)|dx=\hat{I},
\]
\begin{equation}
\langle q(x)|q(x\prime)\rangle=\langle p(x)|p(x\prime)\rangle=\delta
(x-x\prime)
\end{equation}
and
\begin{equation}
\langle q(x)|p(x\prime)\rangle=\dfrac{e^{ixx\prime}}{\sqrt{2\pi}}.
\end{equation}

An\textit{\ ideal} von-Neumann measurement can be defined as an
\textit{instantaneous} interaction between the two subsystems as modeled by
the following delta-like time-pulse hamiltonian operator at time $t_{0}$:
\begin{equation}
\hat{H}_{int}(t)=\lambda\delta(t-t_{0})\hat{O}\otimes\hat{P},
\end{equation}
where $\lambda$ is a parameter that represents the intensity of the
interaction. This ideal situation models a setup where we are supposing that
the time of interaction is very small compared to the time evolution given by
the free Hamiltonians of both subsystems.

Let the initial state of the total system be given by the following
unentangled product state: $|\psi_{i}\rangle=|\alpha\rangle\otimes|\varphi
_{i}\rangle$ and the final state given by $|\psi_{f}\rangle=\hat{U}%
(t_{A},t_{B})|\psi_{i}\rangle\quad(t_{A}<t_{0}<t_{B})$, where the total
unitary evolution operator is%
\begin{equation}
\hat{U}(t_{A},t_{B})=e^{-i\int_{t_{A}}^{t_{B}}\hat{H}_{int}(t)dt}%
=e^{-i\lambda\hat{O}\otimes\hat{P}},
\end{equation}
such that
\begin{equation}
(\hat{I}\otimes\langle q(x)|)|\psi_{f}\rangle=|o_{j}\rangle\otimes\langle
q(x)|\hat{V}_{o_{j}}^{\dagger}|\varphi_{i}\rangle\alpha^{j},
\end{equation}
where $|\alpha\rangle=|o_{j}\rangle\langle o^{j}|\alpha\rangle=|o_{j}%
\rangle\alpha^{j}$ and $\hat{V}_{\xi}$ is the one-parameter family of unitary
operators in $W_{M}$\ that implements the \textit{abelian group of
translations} in the position basis ($x,\xi\in%
\mathbb{R}
$) as $\hat{V}_{\xi}|q(x)\rangle=|q(x-\xi)\rangle$. A correlation in the final
state of the total system is then established between the variable to be
measured $o_{j}$ with the continuous position variable of the measuring
particle:
\begin{equation}
(\hat{I}\otimes\langle q(x)|)|\psi_{f}\rangle=|o_{j}\rangle\alpha^{j}%
\varphi_{i}(x-\lambda o_{j}),
\end{equation}
where $\varphi_{i}(x)=\langle q(x)|\varphi_{i}\rangle$ is the wave-function in
the position basis of the measuring system (the 1-D particle) in its initial
state. This step of the von Neumann measurement prescription is called the
\textit{pre-measurement} of the system.

\section{A discrete measuring system}

Let us consider now the measuring system as a \textit{finite dimensional}
quantum system $W_{M}^{(n)}$. In particular, if $n=2$, our measuring apparatus
consists of a \textit{single} qubit. We shall then start by initially treating
this two-level measuring system so that we may make explicit use of Bloch
sphere geometry and afterwards we shall extend this geometric treatment to
infinite dimensional spaces.

\subsection{Geometry of the space of rays.}

Let $W^{n+1}$ be a $(n+1)$-dimensional Hilbert space together with its dual
$\overline{W}^{n+1}$ and let also $\{|u_{\sigma}\rangle\}$ $(\sigma
=0,1,...,n)$ be an arbitrary basis for $W^{n+1}.$ An hermitean inner product
may be introduced by an \textit{ant-linear} mapping $\dag:W^{n+1}%
\longrightarrow\overline{W}^{n+1}$ (where $\dag$ is the familiar ``dagger"
operation). Indeed, the inner product between two arbitrary states
$|\psi\rangle$ and $|\varphi\rangle$ can now be defined as%
\[
(|\psi\rangle,|\varphi\rangle)=|\psi\rangle\left(  ^{\dagger}|\varphi
\rangle\right)  =\langle\psi|\varphi\rangle.
\]

Thus, an arbitrary normalized ket $|\psi\rangle$ expanded in such a basis can
be represented by a complex $(n+1)$-column matrix:
\begin{equation}
|\psi\rangle=|u_{\sigma}\rangle\psi^{\sigma}\equiv\left(  \psi_{0}\psi
_{1}...\psi_{n}\right)  ^{\intercal}, \quad\text{with\quad}\overline{\psi
}_{\sigma}\psi^{\sigma}=1. \label{column matrix expansion of normalized ket}%
\end{equation}

Writing the complex amplitudes as $\psi^{\sigma}=x^{\sigma}+iy^{\sigma}$ one
can easily see that the set of normalized states can be identified with a
$(2n+1)$-dimensional sphere $S^{2n+1}\subset\mathbb{C}^{n+1}$. Since two state
vectors that differ by a complex phase cannot be physically distinguished by
any means, it is convenient to define the true physical space of states as the
above defined set of normalized states \textit{modulo} the equivalence
relation in $S^{2n+1}$ defined as
\[
|\psi\rangle\sim|\varphi\rangle\iff\exists\quad\theta\in\mathbb{R}
\ /\ |\psi\rangle=e^{i\theta}|\varphi\rangle.
\]

The space of rays defined above is also known as the $n$-dimensional (complex)
projective space $\mathbb{CP}(n)$. A standard complex coordinate system for
$\mathbb{CP}(n)$ is provided by $n$ complex numbers $\xi^{i}=\psi^{i}%
\diagup\psi^{0}$ ($i=1,...,n$) for those points where $\psi^{0}\neq0$. In the
$n=1$ case we have a \textit{single} qubit described by a single complex
coordinate $\xi$. In this case, $\mathbb{CP}(1)$ is topologically equivalent
to a $2D$ sphere and the stereographic projection map $\xi=\tan(\theta
/2)e^{i\varphi}$ provides the Bloch sphere with standard coordinates. Thus,
any\ physical state can be expressed as a normalized state represented as a
point on the Bloch sphere in the following standard form
\begin{equation}
|\psi\rangle=|\theta,\varphi\rangle=\cos\left(  \theta/2\right)  |u_{0}%
\rangle+e^{i\varphi}\sin\left(  \theta/2\right)  |u_{1}\rangle,
\label{qubit on Bloch sphere}%
\end{equation}
where one can easily see that antipode points in the Bloch sphere represent
orthogonal state vectors. In the concluding chapter, we shall see that the
complex number $\xi=\tan(\theta/2)e^{i\varphi}$ can be directly physically
measured as a certain appropriate weak value for two level systems.

\subsection{The pre-measuring interaction}

Suppose now that the interaction happens in $W=W_{S}\otimes W_{M}^{(m)}$ where
the dimension of the measuring system is \textit{finite}:
\[
\dim W_{M}^{(m)}=m.
\]
The initial separable pure-state is $|\psi_{(i)}\rangle=|\alpha\rangle
\otimes|\varphi_{(i)}\rangle$ and $\{|v_{\sigma}\rangle\}$ $(\sigma
=0,1,...,m-1)$ is the finite momentum basis of $W_{M}^{(m)}$ so the momentum
observable can be expressed as $\hat{P}=|v_{\sigma}\rangle p_{\sigma}\langle
v^{\sigma}|$. As in the first section, we model our instantaneous interaction
with the hamiltonian $\hat{H}=\lambda\delta(t-t_{0})\hat{O}\otimes\hat{P}$, so
that for $t_{f}>t_{0}>t_{i}$ one has:
\begin{equation}
|\psi_{(f)}\rangle=\hat{U}(t_{i},t_{f})|\psi_{(i)}\rangle=e^{-i\lambda
p_{\sigma}\hat{O}}|\alpha\rangle\otimes|v_{\sigma}\rangle\varphi^{\sigma},
\end{equation}
where we have expanded $|\varphi_{(i)}\rangle\in W_{M}^{(m)}$ in the finite
momentum basis $\{|v_{\sigma}\rangle\}$. We can now define%
\begin{equation}
|A_{\sigma}\rangle=e^{-i\lambda p_{\sigma}\hat{O}}|\alpha\rangle.
\end{equation}
So that the final state of the overall system at $t_{f}$ will be:
\begin{equation}
|\psi_{(f)}\rangle=|A_{\sigma}\rangle\otimes|v_{\sigma}\rangle\varphi^{\sigma
}. \label{final indexed state}%
\end{equation}
The above entangled state clearly establishes a finite index correlation
between $|A_{\sigma}\rangle$ $\in$ $W_{S}$ and the finite momentum basis
$|v_{\sigma}\rangle$. The total system is in the pure state $|\psi
_{(f)}\rangle\langle\psi_{(f)}|$ and by tracing out the first subsystem, the
measuring system will be:%
\begin{equation}
\hat{\rho}_{|\psi_{(f)}\rangle}^{(m)}=|v_{\sigma}\rangle\varphi^{\sigma
}\langle A^{\tau}|A_{\sigma}\rangle\bar{\varphi}_{\tau}\langle v^{\tau}|.
\label{system density matrix 2}%
\end{equation}

Following Tamate \textit{et al}, we consider the second subsystem (the
measuring system) as a \textit{single qubit}. In this case one may define%
\[
|\varphi_{(i)}\rangle=\cos(\theta/2)|v_{0}\rangle+\sin(\theta/2)e^{i\varphi
}|v_{1}\rangle,
\]
with
\[
\langle A^{0}|A_{1}\rangle=|\langle A^{0}|A_{1}\rangle|e^{-i\beta},
\]
so that we can compute the probability $p(\beta)$ of finding the second
subsystem in a reference state $|\theta=\pi/2,\varphi=0\rangle$ as%
\begin{equation}%
\begin{split}
p(\beta)  &  =tr\left(  \hat{\rho}_{|\psi_{(f)}\rangle}^{(m)}|\pi
/2,0\rangle\left\langle \pi/2,0\right\vert \right) \\
&  =\frac{1}{2}+\frac{1}{4}|\langle A^{0}|A_{1}\rangle|\sin{\theta}%
\cos(\varphi-\beta).
\end{split}
\end{equation}
For a fixed angle $\theta$, this probability is \textit{maximized} when
$\varphi=\beta$. This fact can be used to measure the so called geometric
phase $\beta=\arg(\langle A^{1}|A_{0}\rangle)$ between the two indexed states
$|A_{0}\rangle$ and $|A_{1}\rangle$ $\in$ $W_{S}$. This definition of a
geometric phase was originally proposed in 1956 by Pancharatnam
\cite{Pancharatnam1956} for optical states and rediscovered by Berry in 1984
\cite{berry1984} in his study of the adiabatic cyclic evolution of quantum
states. In 1987, Anandan and Aharonov \cite{aharonov1987} gave a description
of this phase in terms of geometric structures of the $U(1)$ fiber-bundle
structure over the space of rays and of the symplectic and Riemannian
structures in the projective space $\mathbb{CP}(n)$ inherited from the
hermitean structure of $W_{S}$.

\subsection{Phase change due to post-selection}

Given $|\psi_{(f)}\rangle$ resulting from the interaction between both
subsystems we post-select a state $|\beta\rangle$ of $W_{S}$. This procedure
induces a phase change as we shall see. The resulting state after
post-selection is clearly
\begin{equation}
|\psi_{(f)}^{p}\rangle=C(|\beta\rangle\langle\beta|\otimes\hat{I})(|A_{\sigma
}\rangle\otimes|v_{\sigma}\rangle\varphi^{\sigma}),
\end{equation}
where $C$ is an unimportant normalization constant. Because of the
post-selection, the system is in a non-entangled state so that the partial
trace of $\hat{\rho}_{|\psi_{(f)}^{p}\rangle}=|\psi_{(f)}^{p}\rangle
\langle\psi_{(f)}^{p}|$ over the first subsystem gives us%
\begin{equation}
\label{final state after post-selection}|\varphi_{(f)}\rangle=C\langle
\beta|A_{\sigma}\rangle\varphi^{\sigma}|v_{\sigma}\rangle.
\end{equation}

Making the following phase choices $\langle\beta|A_{0}\rangle=|\langle
\beta|A_{0}\rangle|e^{i\beta_{0}}$ and $\langle\beta|A_{1}\rangle
=|\langle\beta|A_{1}\rangle|e^{-i\beta_{1}}$, we can again compute the
probability of finding the second subsystem in state $|\pi/2,0\rangle$:
\[%
\begin{split}
p  &  =\frac{C^{2}}{2}[|\langle\beta|A_{0}\rangle|^{2}\cos^{2}(\theta
/2)+|\langle\beta|A_{1}\rangle|^{2}\sin^{2}(\theta/2)+\\
&  + \sin\theta|\langle\beta|A_{0}\rangle\langle\beta|A_{1}\rangle
|\cos(\varphi_{p}-\beta_{0}-\beta_{1})]
\end{split}
\]
For a fixed angle $\theta$, the maximum probability occurs for $\varphi
_{p}=\beta_{0}+\beta_{1}=\arg(\langle\beta|A_{0}\rangle\langle A^{1}%
|\beta\rangle)$. This implies that there is an overall phase change $\Theta$
given by
\begin{equation}
\label{geometric invariant}\Theta=\varphi_{p}-\varphi=\arg(\langle A^{1}%
|\beta\rangle\langle\beta|A_{0}\rangle\langle A^{0}|A_{1}\rangle) .
\end{equation}

The quantity given by (\ref{geometric invariant}) is a geometric invariant in
the sense that it depends only on the projection of the state vectors
$|A_{0}\rangle$, $|A_{1}\rangle$ and $|\beta\rangle$ on $\mathbb{CP}(n)$. In
fact, this quantity is the intrinsic geometric phase picked by a state vector
that is parallel transported through the closed geodesic triangle defined by
the projection of the three states on ray space.

For a single qubit, the geometric invariant is proportional to the area of the
geodesic triangle formed by the projection of the kets ($|A_{0}\rangle$,
$|A_{1}\rangle$ and $|\beta\rangle$) on Bloch sphere and it is well known to
be given by
\begin{equation}
\Theta=\arg(\langle A^{0}|\beta\rangle\langle\beta|A_{1}\rangle\langle
A^{1}|A_{0}\rangle)=-\frac{\Omega}{2}, \label{geometric invariant2}%
\end{equation}
where $\Omega$ is the oriented solid angle formed by the geodesic triangle.
\begin{figure}[ptb]
\begin{center}
\includegraphics[height=2.5in,width=2.5in]{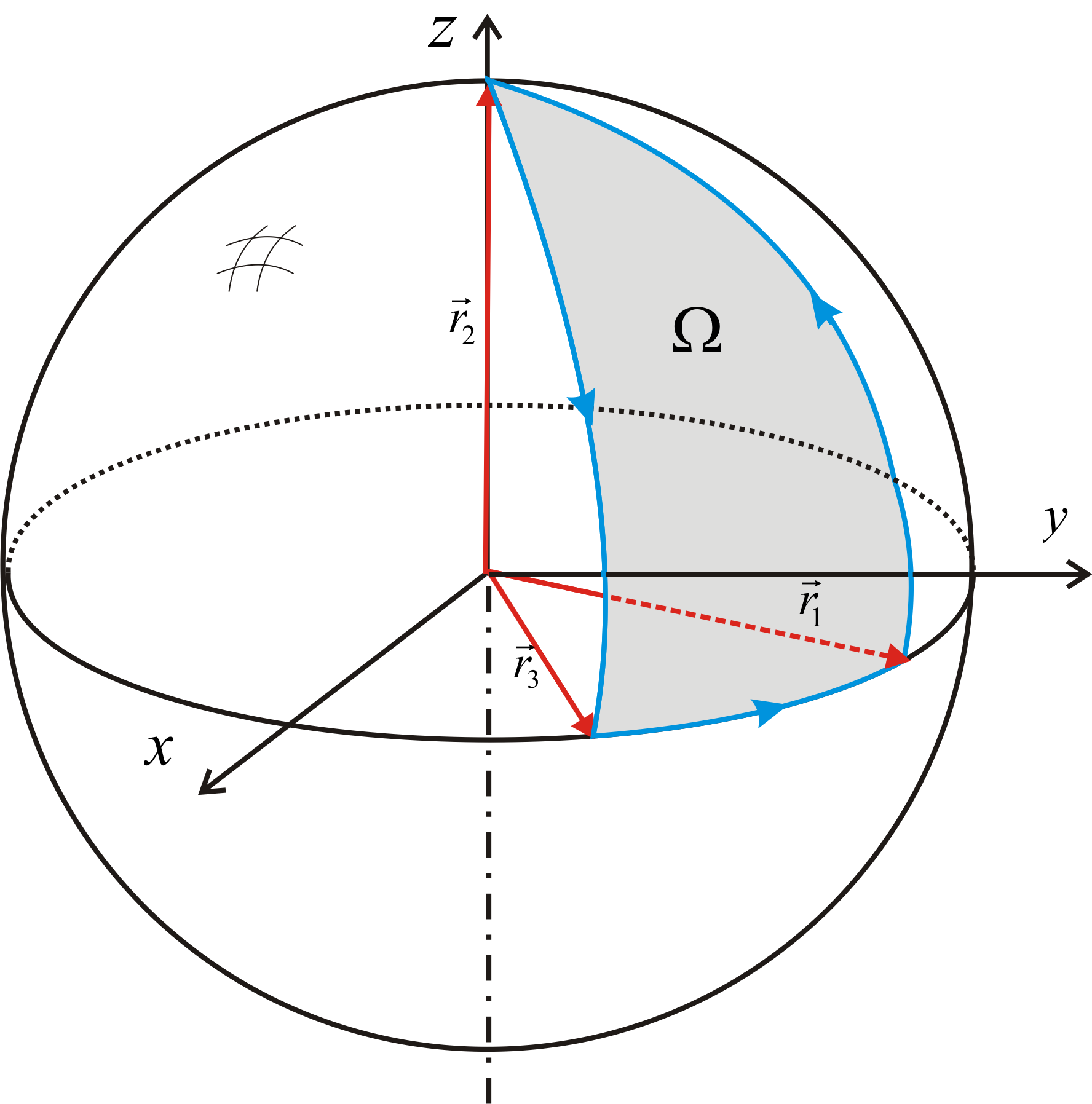}
\end{center}
\caption{Solid angle determined by 3 points: the north pole and 2 points on
the equator of the Bloch sphere.}%
\end{figure}

\section{The measuring system with a continuous base.}

Suppose a physical system $W$ is composed by two subsystems $W_{S}\otimes
W_{M}^{(\infty)}$ as before, but the measuring system $W_{M}^{(\infty)}$ is
spanned \ by complete sets of position kets $\{|q(x)\rangle\}$ (momentum kets
$\{|p(y)\rangle\}$), with $-\infty<x,y<+\infty$. Let us consider $|\psi
_{(i)}\rangle=|\alpha\rangle\otimes|\varphi_{(i)}\rangle$ as the initial
product state and $\hat{H}=\lambda\delta(t-t_{0})\hat{O}\otimes\hat{P}$, with
$\hat{P}=\int_{-\infty}^{+\infty}y|p(y)\rangle\langle p(y)|dy$, the
hamiltonian that models the instantaneous measuring interaction so that the
system evolves to%
\begin{equation}
|\psi_{(f)}\rangle=\hat{U}(t_{i},t_{f})|\psi_{(i)}\rangle=\int_{-\infty
}^{+\infty}dye^{-i\lambda y\hat{O}}|\alpha\rangle\otimes|p(y)\rangle
\varphi_{p}(y),
\end{equation}
where $\varphi_{p}(y)=\langle p(y)|\varphi_{(i)}\rangle$ is the momentum wave
function associated to state $|\varphi_{(i)}\rangle$. We may define the state
\begin{equation}
|A(y)\rangle=e^{-i\lambda y\hat{O}}|\alpha\rangle,
\label{continuous indexed state os the measured system}%
\end{equation}
so that we can rewrite the ket $|\psi_{(f)}\rangle$ as
\begin{equation}
\label{final continously indexed state}|\psi_{(f)}\rangle=\int_{-\infty
}^{+\infty}dy|A(y)\rangle\otimes|p(y)\rangle\varphi_{p}(y),
\end{equation}
where the states $|A(y)\rangle$ are indexed by the continuous parameter $y\in%
\mathbb{R}
$. We may now compute (to first order in $dy$) the intrinsic phase shift
between $|A(y)\rangle$ and $|A(y+dy)\rangle$ in a similar way that was carried
out in the previous section with the discretely parametrized states:%
\begin{equation}
\arg(\langle A(y)|A(y+dy)\rangle)\approx-\lambda dy\langle\hat{O}\rangle_{|
\alpha\rangle}, \label{phase shift for the continuum}%
\end{equation}
where $\langle\hat{O}\rangle_{|\alpha\rangle}=\langle\alpha|\hat{O}%
|\alpha\rangle$ is the expectation value of observable $\hat{O}$ in state
$|\alpha\rangle$.

We can also compute the shift of the expectation value of the position
observable $\hat{Q}$ of the particle of the measuring system between the
initial and final states. Let $\{|o_{j}\rangle\},$ ($j=0,...,N-1$) be a
complete set of eigenkets of observable $\hat{O}$. The final state of the
composite system can be described by the following pure density matrix:%
\begin{equation}
\hat{\rho}_{|\psi_{(f)}\rangle}=|\psi_{(f)}\rangle\langle\psi_{(f)}%
|=|o_{j}\rangle\langle o^{k}|\otimes\alpha^{j}\hat{V}_{\lambda o_{j}}%
^{\dagger}|\varphi_{(i)}\rangle\langle\varphi_{(i)}|\hat{V}_{\lambda o_{k}%
}\bar{\alpha}_{k}.
\end{equation}
Taking the partial trace of the $W_{S}$ system, we arrive at the following
mixed state that describes the measuring system at instant $t_{f}$:
\begin{equation}
\hat{\rho}_{|\psi_{(f)}\rangle}^{(M)}=\sum_{j}|\alpha^{j}|^{2}\hat{V}_{\lambda
o_{j}}^{\dagger}|\varphi_{(i)}\rangle\langle\varphi_{(i)}|\hat{V}_{\lambda
o_{j}}.
\end{equation}
The ensemble expectation value $[\hat{Q}]_{\hat{\rho}_{|\psi_{(f)\rangle}%
}^{(M)}}$ of position is then given by:
\begin{equation}
\lbrack\hat{Q}]_{\hat{\rho}_{|\psi_{(f)}\rangle}^{(M)}}=tr(\hat{\rho}%
_{|\psi_{(f)}\rangle}^{(M)}\hat{Q})=\langle\hat{Q}\rangle_{|\varphi
_{(i)}\rangle}+\lambda\langle\hat{O}\rangle_{|\alpha\rangle}.
\label{shift in position}%
\end{equation}

The above result is similar to the one obtained by Tamate \textit{et al}, yet
we believe that the procedure we have adopted is mathematical more precise as
we will discuss in the final concluding section of this paper. One may ask at
this point if a similar procedure may be carried out in the case of weak
values, since these can be thought of as a generalization of expectation
values. The answer is affirmative, but before we demonstrate this, we shall
discuss in the next section, a geometrical interpretation also inspired by
Tamate \textit{et al's} description of the interaction between the system
$W_{S}$ and the measuring system.

\subsection{Geometric interpretation of von Neumann's pre-measurement}

Let $W^{n+1}$ be a $(n+1)$-dimensional Hilbert space with basis $\{|u_{\sigma
}\rangle\}$ so that an arbitrary (not necessarily normalized) vector of this
space is described as $|\psi\rangle=|u_{\sigma}\rangle\psi^{\sigma}$, where
greek indices take values in $\sigma=0,...,n$. One can map this state to a
sphere $S^{2n+1}$ with radius given by
\begin{equation}
\bar{\psi}_{\sigma}\psi^{\sigma}=r^{2}. \label{unnormalized vector}%
\end{equation}
We introduce \textit{projective coordinates} $\xi^{i}$ on $\mathbb{CP}(n)$ so
that
\begin{equation}
\psi^{0}=\frac{re^{i\varphi}}{(1+\bar{\xi}_{i}\xi^{i})^{1/2}},\quad
\text{with}\quad i=1,...,n, \label{psi 0}%
\end{equation}
where $\varphi$ is an arbitrary phase factor. The \textit{euclidean metric} in
$W^{n+1}$, seen here as a $(2n+2)$-dimensional \textit{real} vector space, can
be written as \cite{page1987}
\begin{equation}
ds^{2}(W^{n+1})=d\psi^{\sigma}d\bar{\psi}_{\sigma}=dr^{2}+r^{2}ds^{2}%
(S^{2n+1}), \label{metric on W(n+1)}%
\end{equation}
where%
\begin{equation}
ds^{2}(S^{2n+1})=(d\varphi-A)^{2}+ds^{2}(\mathbb{CP}(n))
\label{metric on the sphere of normalized states}%
\end{equation}
is the squared distance element over the space of normalized vectors, the
$(2n+1)$-sphere, in $W^{n+1}$ and
\begin{equation}
A=\frac{i}{2}\left(  \frac{\xi^{i}d\bar{\xi}_{i}-\bar{\xi}_{i}d{\xi}^{i}%
}{1+\bar{\xi}_{i}{\xi}^{i}}\right)  \label{abelian conection 1-form}%
\end{equation}
is the well known abelian 1-form connection of the $U(1)$ bundle over
$\mathbb{CP}(n)$ and $ds^{2}$ is the metric over $\mathbb{CP}(n)$ in
projective coordinates given explicitly by: \cite{page1987,chruscinski2004}
\begin{equation}
ds^{2}(\mathbb{CP}(n))=\left[  \frac{(1+\bar{\xi}_{i}{\xi}^{i})\delta_{j}%
^{k}-\bar{\xi}_{k}\xi^{j}}{(1+\bar{\xi}_{i}{\xi}^{i})^{2}}\right]  d\xi
^{k}d\bar{\xi}_{j}. \label{metric on CP(n)}%
\end{equation}

\begin{figure}[th]
\begin{center}
\includegraphics[
height=2.5284in,
width=2.507in]{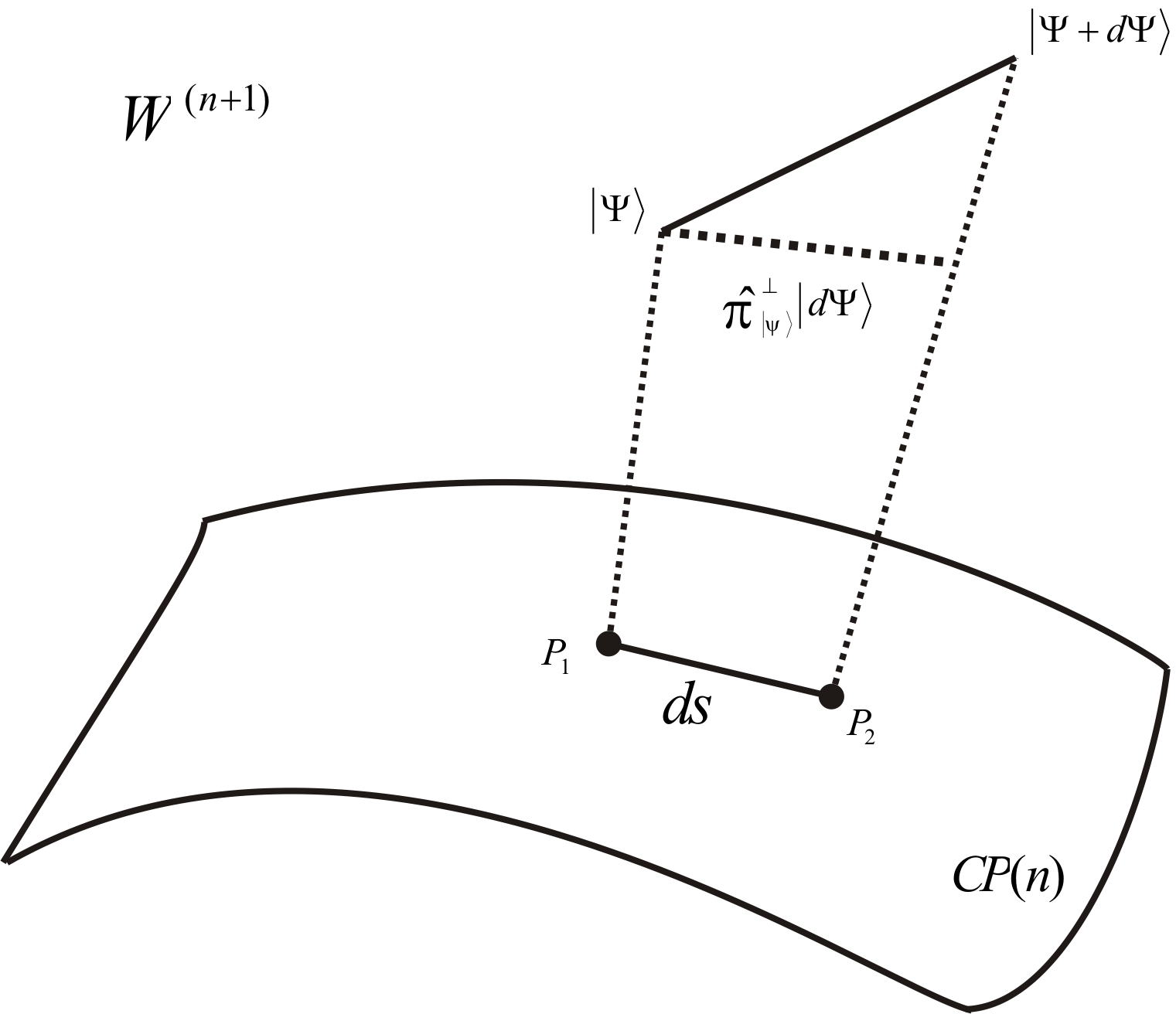}
\end{center}
\caption{Pictorial representation of the quantum space of states}%
\label{figura_espaco}%
\end{figure}

A natural and intuitive picture of these structures can be seen easily in FIG.
\ref{figura_espaco}. The points $P_{1}$ and $P_{2}$ $\in$ $\mathbb{CP}(n)$ are the
projections respectively from two infinitesimally nearby normalized state
vectors $|\psi\rangle$ and $|\psi+d\psi\rangle$. It is natural to define then,
the squared distance between $P_{1}$ and $P_{2}$ as the projection of
$|d\psi\rangle$ in the\ ``orthogonal direction" of $|\psi\rangle$, that is,
the projection given by the projection operator $\hat{\pi}_{|\psi\rangle
}^{\perp}=\hat{I}-|\psi\rangle\langle\psi|$ as shown in \ref{figura_espaco}.
It is then easy to see that
\begin{equation}
ds^{2}(\mathbb{CP}(n))=\langle d\psi|d\psi\rangle-\langle d\psi|\psi
\rangle\langle\psi|d\psi\rangle. \label{metric 2}%
\end{equation}
The above equation is an elegant manner to express (\ref{metric on CP(n)}). By
inspecting both (\ref{metric on the sphere of normalized states}) and
(\ref{metric 2}), it is not difficult to conclude that
\begin{equation}
(d\varphi-A)^{2}=\langle d\psi|\psi\rangle\langle\psi|d\psi\rangle.
\end{equation}

Let $|\psi(t)\rangle$ be the curve of normalized state vectors in $W^{n+1}$
given by the unitary evolution generated by an hamiltonian $\hat{H}$. The
Schr\"{o}dinger equation implies a relation between $|\psi(t)\rangle$ and
$|\psi(t+dt)\rangle$ given by:
\begin{equation}
|d\psi\rangle=|\psi(t+dt)\rangle-|\psi(t)\rangle=-i\hat{H}|\psi(t)\rangle dt.
\label{infinitesimal unitary displacement}%
\end{equation}
The above equation together with (\ref{metric 2}) lead to a very elegant
relation for the squared distance between two infinitesimally nearby
projection of state vectors connected by the unitary evolution over
$\mathbb{CP}(n)$ \cite{Anandan1990}:
\begin{equation}%
\begin{split}
ds^{2}(\mathbb{CP}(n))  &  =\left[  \langle\psi(t)|\hat{H}^{2}|\psi
(t)\rangle-(\langle\psi(t)|\hat{H}|\psi(t)\rangle)^{2}\right]  dt^{2}\\
&  =\left(  \delta_{|\psi(t)\rangle}^{2}E\right)  dt^{2}.\label{metric 3}%
\end{split}
\end{equation}
One may say that the equation above means that the speed of the projection
over $\mathbb{CP}(n)$ equals the instantaneous energy uncertainty
\begin{equation}
\frac{ds}{dt}=\delta E(t). \label{speed over CP(n)}%
\end{equation}

A beautiful geometric derivation of the \textit{time-energy uncertainty
relation} that follows directly from (\ref{speed over CP(n)}) can be found in
\cite{Anandan1990}. Back to our discussion of the interaction between the
systems $W_{S}$ and $W_{M}^{(\infty)}$, note that equation
(\ref{continuous indexed state os the measured system}) is formally equivalent
to the unitary time evolution equation $|\psi(t)\rangle=e^{-i\hat{H}t}%
|\psi(0)\rangle$ which is clearly a solution of a Schr\"{o}dinger equation
with \textit{time-independent} hamiltonian. A formal analogy between the two
distinct physical processes is exemplified by the association below:
\[
\begin{array}
[c]{ll}%
|\psi(t)\rangle & \mapsto|A(y)\rangle\\
|\psi(0)\rangle & \mapsto|\alpha\rangle=|A(0)\rangle\\
t & \mapsto y\\
\hat{H} & \mapsto\lambda\hat{O}.
\end{array}
\]

\begin{figure}[ptb]
\begin{center}
\includegraphics[height=2.5in,width=2.5in]{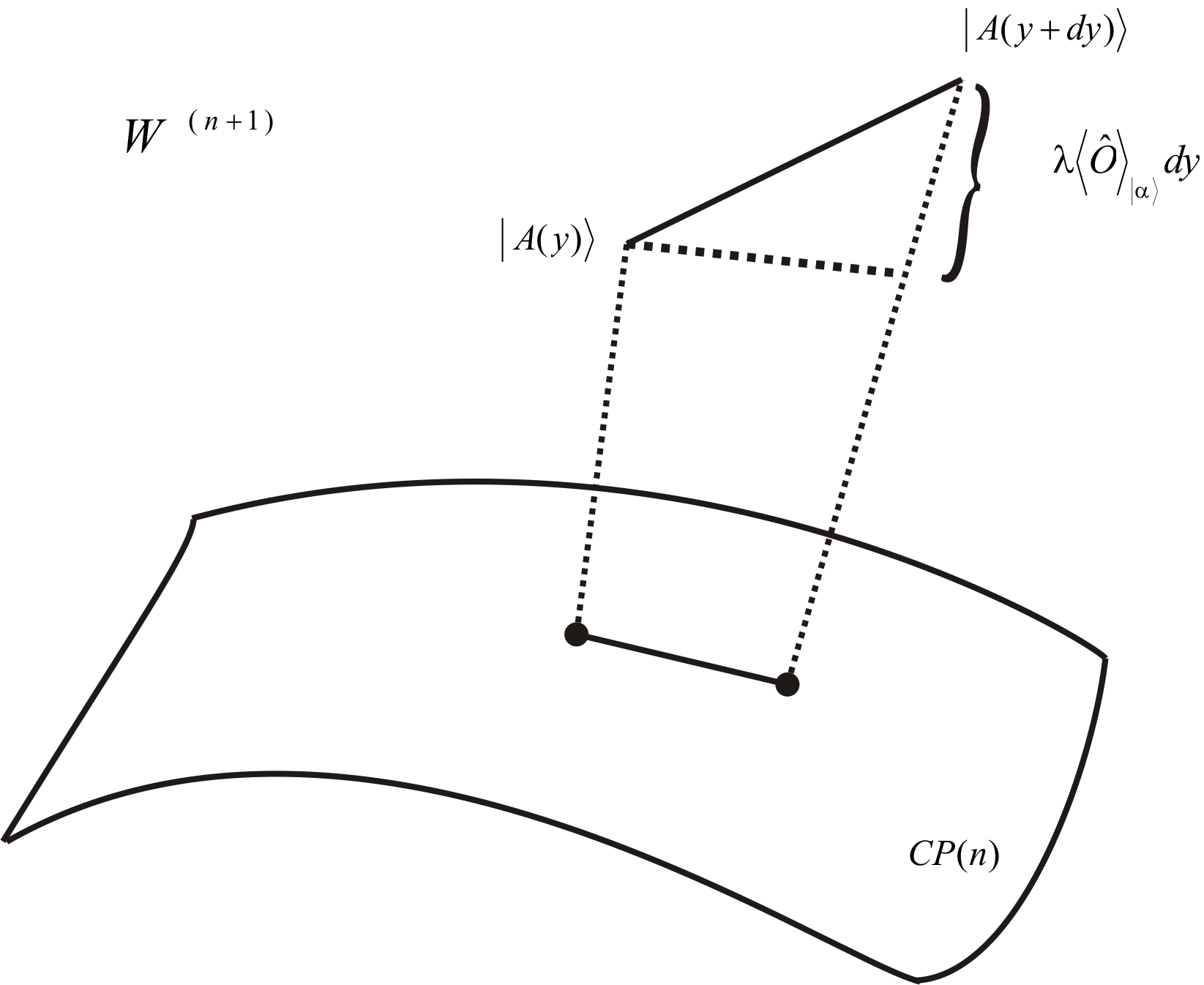}
\end{center}
\caption{Pictorial representation of the phase difference between
$|A(y)\rangle$ and $|A(y+dy)\rangle$.}%
\label{figura_espaco2}%
\end{figure}

Looking at subsystem $W_{S}$ and regarding $y$ as an external parameter (just
like the time variable for the unitary time evolution) we may write the analog
of (\ref{metric 3}) in $\mathbb{CP}(n)\subset W_{S}$:
\begin{equation}
\label{line element over CP(n) with parameter y}%
\begin{split}
ds^{2}  &  =\left[  \langle A(y)|\hat{O}^{2}|A(y)\rangle-\langle A(y)|\hat
{O}|A(y)\rangle^{2}\right]  \lambda^{2}dy^{2}\\
&  =\left[  \langle\alpha|\hat{O}^{2}|\alpha\rangle-\langle\alpha|\hat
{O}|\alpha\rangle^{2}\right]  \lambda^{2}dy^{2}.
\end{split}
\end{equation}
Comparing this result with (\ref{phase shift for the continuum}) and
(\ref{metric 2}), we advance one step further than Tamate \textit{et al} in
their geometrization programme as we present a geometric interpretation for
the expectation value $\langle\alpha|\hat{O}|\alpha\rangle$ in terms of the
$U(1)$ fiber bundle structure as one can easily infer from the pictorial
representation in FIG. \ref{figura_espaco2}. 

\subsection{Post-selection and weak values}

\begin{figure}[ptb]
\begin{center}
\includegraphics[
height=2.5284in,
width=2.507in]{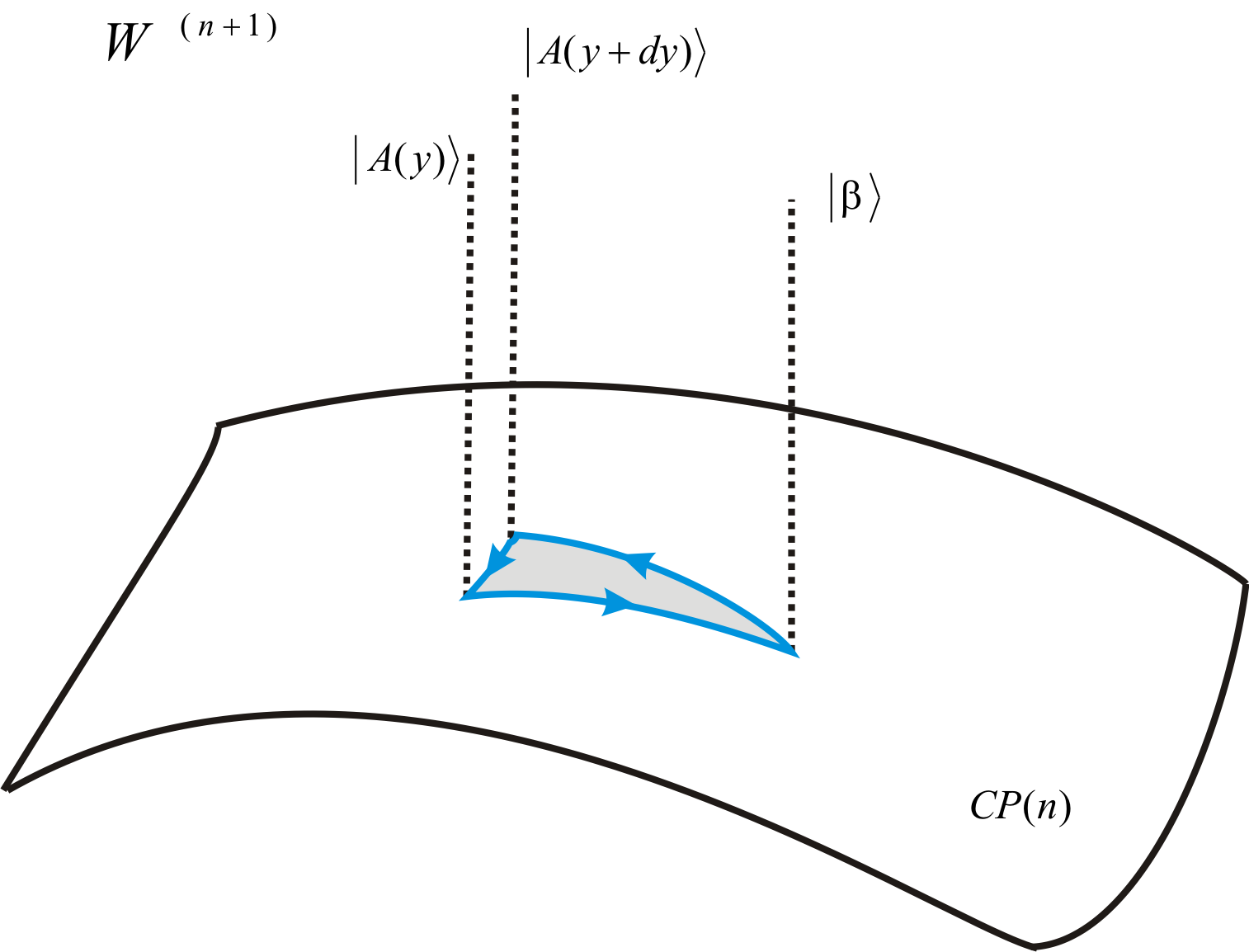}
\end{center}
\caption{Pictorial representation of the global geometric phase.}%
\label{areavarrida}%
\end{figure}

For the case of a \textit{weak measurement}, the hamiltonian can be modeled as
$\hat{H}^{(w)}=\epsilon\delta(t-t_{0})\hat{O}\otimes\hat{P}$, with
$\epsilon\rightarrow0$ \cite{aharonov1988}. Given the initial unentangled
state $|\psi_{(i)}\rangle=|\alpha\rangle\otimes|\varphi_{(i)}\rangle$ at
$t_{0}$, such that $t_{i}<t_{0}<t_{f}$, the system is described as%
\begin{align*}
|\psi_{(f)}\rangle &  =\hat{U}(t_{i},t_{f})|\psi_{(i)}\rangle=e^{-i\epsilon
\hat{O}\otimes\hat{P}}|\alpha\rangle\otimes|\varphi_{(i)}\rangle\\
&  =\int_{-\infty}^{+\infty}dy|A(y)\rangle\otimes|p(y)\rangle\varphi_{p}(y),
\end{align*}
with $|A(y)\rangle=e^{-i\epsilon y\hat{O}}|\alpha\rangle$. The global
geometric phase related to the infinitesimal geodesic triangle formed by the
projections of $|A(y)\rangle$, $|A(y+dy)\rangle$ and the post-selected state
$|\beta\rangle$ on $\mathbb{CP}(n)$ (see FIG. \ref{areavarrida}) is given by:%
\[
\Theta=\arg\left[  \langle A(y)|\beta\rangle\langle\beta|A(y+dy)\rangle\langle
A(y+dy)|A(y)\rangle\right]  .
\]
Expanding to first order in $\epsilon$, we finally obtain%
\begin{equation}
\Theta\approx-\epsilon\left[  \operatorname{Re}({O}_{w})-\langle\hat{O}%
\rangle_{|\alpha\rangle}\right]  dy,
\end{equation}
where $O_{w}=\langle\beta|\hat{O}|\alpha\rangle/\langle\beta|\alpha\rangle$ is
the weak value of $\hat{O}$ and $\langle\hat{O}\rangle_{|\alpha\rangle}$ is
the expectation value of $\hat{O}$ in state $|\alpha\rangle$. Following the
same approach of section III, we can compute the expectation value of the
position observable $\hat{Q}$ of the measuring system $W_{M}^{(\infty)}$
between the initial and final states. The final state after post-selection of
a state $|\beta\rangle$ of system $W_{S}$ is given by%
\begin{align*}
|\psi_{(f)}\rangle &  =C(|\beta\rangle\langle\beta|\otimes\hat{I}%
)(e^{-i\epsilon\hat{O}\otimes\hat{P}}|\alpha\rangle\otimes|\varphi
_{(i)}\rangle)\\
&  \approx C(|\beta\rangle\langle\beta|\otimes\hat{I})(\hat{I}-i\epsilon
\hat{O}\otimes\hat{P})|\alpha\rangle\otimes|\varphi_{\left(  i\right)
}\rangle,
\end{align*}
where $C\approx(1+\epsilon\langle\hat{P}\rangle_{|\alpha\rangle}%
\operatorname{Im}({O}_{w})) \diagup\langle\beta|\alpha\rangle$ is the
normalization constant because, in general, the state after post-selection is
\textit{not} normalized. By partial tracing out the first subsystem we arrive
at:%
\[%
\begin{split}
\hat{\rho}_{|\psi_{(f)}\rangle}^{(2)}  &  =tr_{1}(|\psi_{(f)}\rangle
\langle\psi_{(f)}|)\\
&  =\left[  1-i\epsilon\langle\hat{P}\rangle|\varphi_{\left(  i\right)
}\rangle({O}_{w}-\bar{O}_{w})\right]  |\varphi_{\left(  i\right)  }%
\rangle\langle\varphi_{\left(  i\right)  }| -\\
&  -i\epsilon({O}_{w}\hat{P}|\varphi_{\left(  i\right)  }\rangle\langle
\varphi_{\left(  i\right)  }|-\bar{O}_{w}|\varphi_{\left(  i\right)  }%
\rangle\langle\varphi_{\left(  i\right)  }|\hat{P}),
\end{split}
\]
where $\langle\hat{P}\rangle_{|\varphi_{(i)}\rangle}$ is the expectation value
of momentum $\hat{P}$ of the measuring system in state $|\varphi_{(i)}\rangle$
and $\bar{{O}}_{w}$ is the complex conjugate of the weak value ${O}_{w}$. The
shift in the ensemble average $[\hat{Q}]_{\hat{\rho}_{|\psi_{(f)}\rangle
}^{(2)}}=tr(\hat{\rho}_{|\psi_{(f)}\rangle}^{(2)}\hat{Q})$ can then be easily
computed as $\Delta\hat{Q}=[\hat{Q}]_{\hat{\rho}_{|\psi_{(f)}\rangle}^{(2)}%
}-[\hat{Q}]_{\hat{\rho}_{|\psi_{(i)}\rangle}^{(2)}}$, giving us
\begin{equation}%
\begin{split}
\Delta\hat{Q}  &  =\epsilon\Big{[}(\operatorname{Im}(O_{w}))(\langle
\varphi_{(i)}|\{\hat{Q},\hat{P}\}|\varphi_{(i)}\rangle-\\
&  -2\langle\hat{P}\rangle_{|\varphi_{(i)}\rangle}\langle\hat{Q}%
\rangle_{|\varphi_{(i)}\rangle}) +\operatorname{Re}(O_{w}%
)\Big{]}.\label{shift in position ensemble average}%
\end{split}
\end{equation}

\section{Concluding Remarks}

In \cite{tamate2009}, the authors introduced a very interesting geometric
interpretation for von Neumann's ideal pre-measurement concept as well as for
the weak value. In this paper we have carried out a review of their paper,
advancing a step further the geometric concepts they introduced in their paper
and clarifying some of their results and calculations. For instance, the
equation (\ref{phase shift for the continuum}) below
\[
\arg(\langle A(y)|A(y+dy)\rangle)\approx-\lambda dy\langle\hat{O}%
\rangle_{|\alpha\rangle}%
\]
is essentially the same result of equation 16 in \cite{tamate2009}:
\begin{equation}
\Theta(y)=\arg(\langle A(0)|A(y)\rangle)\approx-\lambda y\langle\hat{O}%
\rangle_{|\alpha\rangle}.
\end{equation}
Yet, our approach seems to be more mathematically precise as it firmly
grounded on the geometrical structures involved. The authors express a
infinitesimal phase shift by differentiating a \textquotedblleft function"
$\Theta(y)$, but no such function exists because the geometric phase is
obtained from the 1-form $A=i\left(  \xi^{i}d\bar{\xi}_{i}-\bar{\xi}_{i}d{\xi
}^{i}\right)  \diagup2\left(  1+\bar{\xi}_{i}{\xi}^{i}\right)  $. The exterior
derivative $F=dA$ measures the local curvature of the connection form which
measures the local lack of holonomy of the process of comparing intrinsic
phases between normalized state vectors. This means that the 1-form $A$ is
\textit{not} the exterior derivative of any scalar function (a 0-form). The
authors introduced this \textquotedblleft function" $\Theta(y)$ and by
formally taking its derivative, they managed to arrive at the correct
equation
\[
\Delta\hat{Q}=\lambda\langle\hat{O}\rangle.
\]
This result is the same we obtained in (\ref{shift in position}), but, from
the discussion above, it is quite clear that our approach seems to be
mathematically more sound. The authors also approach a geometric
interpretation of weak values, where they found the following equation for the
shift in the expectation value of the position observable (equation 21 of
\cite{tamate2009}):%
\[
\Delta\hat{Q}=\epsilon\operatorname{Re}(O_{w}).
\]
Yet it is well known that this result can be extended to a full complex-valued
weak value (see \cite{jozsa2007} and \cite{lobo2009weak}). The above equation
lacks a term proportional to the\textit{ imaginary part} of the weak value
$O_{w}$ as one can see from equation (\ref{shift in position ensemble average}%
). In fact, in their paper, they calculated an example for a qubit as the
measuring system where they have chosen a very particular set of pre and
post-selected states and observable that assures a weak value with null
imaginary part. Indeed, if we choose the following: $\left\vert \alpha
\right\rangle =\left\vert u_{0}\right\rangle $ (the \textquotedblleft north
pole" of the Bloch sphere), $\left\vert \beta\right\rangle =|\theta
,\varphi\rangle=\cos\left(  \theta/2\right)  |u_{0}\rangle+e^{i\varphi}%
\sin\left(  \theta/2\right)  |u_{1}\rangle$ as respectively the pre and
post-selected states and $\hat{O}=\hat{\sigma}_{1}=|u_{0}\rangle\left\langle
u^{1}\right\vert +|u_{1}\rangle\left\langle u^{0}\right\vert $ as the
observable, then it is straightforward to compute the weak value as
$O_{w}=\tan\left(  \theta/2\right)  e^{i\varphi}$ which is clearly
complex-valued in general. Yet, the post-selected $\left\vert \beta
\right\rangle $ state chosen in \cite{tamate2009} is equivalent to our choice
with the phase $\varphi=0$. This is an \textit{arbitrary restriction} over all
possible choices of states in the Bloch sphere and only for $\varphi=0$ and
$\varphi=\pi$ one arrives at a purely \textit{real} weak value. What is
curious about this result (for a single qubit) is that the weak value gives a
direct physical meaning to the complex projective coordinate $\xi=\tan\left(
\theta/2\right)  e^{i\varphi}$. Indeed, when the experimentalist measures the
(complete complex) weak value of a two level system in his lab, he actually is
directly measuring the point on the $\mathbb{CP}(1)$ (complex plane + a point
in infinity) of $\left\vert \beta\right\rangle $ related to the Bloch sphere
by the stereographic projection. If the post-selected state is somewhere near
the south pole, it is expected that there should be large measured distortions
because of the nature of the projection. To remedy this, it is enough to
rotate $\left\vert \alpha\right\rangle $ and $\hat{O}$ appropriately so that
one can cover all states in $\mathbb{CP}(1)$ with good precision. It would be
interesting to pursue further this kind of investigation of the geometrical
meaning of weak values for higher dimensional systems. For instance, for
higher spin systems, the geometry of spin coherent states could be useful for
this purpose \cite{perelomov1986generalized}. In a preliminary version of our
manuscript we have had the chance to see a reply of Tamate and collaborators
to our paper. In their short reply they manage to explain further why they
have restricted their attention only to the \textit{real part} of the weak
value. It became clear to us that the term
\[
C(\hat{Q},\hat{P})=\langle\varphi_{(i)}|\{\hat{Q},\hat{P}\}|\varphi
_{(i)}\rangle-2\langle\hat{P}\rangle_{|\varphi_{(i)}\rangle}\langle\hat
{Q}\rangle_{|\varphi_{(i)}\rangle}%
\]
in equation (\ref{shift in position ensemble average}) is expected to vanish
for most experimental implementations. This is because for the usual initial
states of the measuring apparatus, the position and momentum observables are
uncorrelated. This is very unfortunate as our example shows that both
imaginary and real parts of the weak value are true elements of reality that
should be treated with the same ontological status. Maybe an experimental
approach that focus on the geometric structures of the phase space of the
measuring apparatus (the pointer) could furnish experimental methods to
accomplish this as we have suggested in \cite{augusto2009}.

The concept of weak values has lately become increasingly important both for
theoretical and experimental reasons \cite{popescu2009}. A deeper
understanding of the physical and the mathematical structures behind weak
values is of urgent need. One possible approach is to look at the phase space
of the measuring system as was carried out in \cite{augusto2009}. Another
promising approach is the one initiated by Tamate\textit{ et al} in
\cite{tamate2009} where they look at the natural geometric structures of the
\textit{measured} system to characterize the weak value concept. We have tried
to continue such geometrization programme by clarifying some conceptions in
their original paper and advancing a step further in this approach. We
introduced a geometric interpretation for the expectation value $\langle
\hat{O}\rangle_{|\alpha\rangle}$ of an arbitrary observable $\hat{O}$ in terms
of the $U(1)$ fiber bundle structure over the projective space of the measured
subspace. We hope that this will lead to further fruitful theoretical and
experimental applications. One possible research path is to consider the
projective space structures of \textit{both} subsystems and try to relate the
exchange of information (in some kind of measure) between them in the (weak or
strong) pre-measurement process in terms of these very same geometric structures.

\section*{Acknowledgments}

A. C. Lobo wishes to acknowledge financial support from
\textit{NUPEC-Funda\c{c}\~{a}o Gorceix} and both authors wish to acknowledge
financial support from \textit{Conselho Nacional de Desenvolvimento
Cient\'{\i}fico e Tecnol\'{o}gico} (CNPq). The authors also thank Tamate and
collaborators for their reply and also thank the assistance with the English
language from Fernanda Lobo Bellehumeur.

{\normalsize
\bibliographystyle{plain}
\bibliography{referencias}
}

\end{document}